\documentclass[12pt,preprint]{aastex}
\shorttitle{ACS imaging of 25 galaxies}
\shortauthors{Karachentsev et al.}
\begin{document}
\title{ ACS imaging of 25 galaxies in nearby groups and in the field}

\author{Igor D.\ Karachentsev}
\affil{Special Astrophysical Observatory, Russian Academy
	  of Sciences, N.\ Arkhyz, KChR, 369167, Russia}
\email{ikar@luna.sao.ru}
\author{Andrew Dolphin}
\affil{Steward Observatory, 933 N. Cherry Ave., Tucson, AZ 85721}
\author{R. Brent Tully}
\affil{University of Hawaii}
\author{Margarita Sharina, Lidia Makarova and Dmitry Makarov}
\affil{Special Astrophysical Observatory, Russian Academy
	  of Sciences, N.\ Arkhyz, KChR, 369167, Russia}
\author{Valentina Karachentseva}
\affil{Astronomical Observatory of Kiev University, Kiev, Ukraine}
\author{Shoko Sakai}
\affil{University of California - Los Angeles USA}
\author{Edward J. Shaya}
\affil{U. of Maryland, Astronomy Dept. - College Park MD, USA}
\begin{abstract}
  We present HST/ACS images and color-magnitude diagrams for 25 nearby
galaxies with radial velocities $V_{LG}< 500$ km s$^{-1}$. 
Distances are determined based on the luminosities of stars at the 
tip of the red giant branch that range from 2~Mpc to 12~Mpc.
Two of the galaxies, NGC~4163 and IC~4662, are found to be the nearest known
representatives of blue compact dwarf (BCD) objects.

  Using high-quality data on distances and radial velocities of  110 nearby
field galaxies, we derive their mean Hubble ratio to be
68 km s$^{-1}$ Mpc$^{-1}$  with standard deviation
of 15 km s$^{-1}$ Mpc$^{-1}$. Peculiar velocities of
most of the galaxies, $V_{pec} = V_{LG} - 68 D$, follow a Gaussian
distribution with $\sigma_v = 63$ km s$^{-1}$, but with a tail
towards high negative values. Our data displays the known
correlation between peculiar velocity and galaxy elevation above the
Local Supercluster plane. The small observed fraction of galaxies with
high peculiar velocities, $V_{pec} < -500$ km s$^{-1}$, may be
understood as objects associated with nearby groups
(Coma I, Eridanus) outside the Local volume.

\end{abstract}

\section{Introduction}

  Until 2000, very little data have been available to describe the peculiar
velocity field of galaxies around the Local Group. This surprising
situation was caused by the lack of reliable data on distances (not
velocities) for many of the nearest galaxies. The Local Group itself is
in the highly nonlinear collapse regime. In a larger volume,
deviations from pure Hubble expansion can be expected due to
the gravitational action of nearby groups as well as by
Virgo-centric and Great Attractor flows. Enormous progress has been
made recently with accurate distance measurements of nearby galaxies
beyond the Local Group based on the luminosity of the tip of the red
giant branch (TRGB). This method (Lee et al. 1993) has
a precision comparable to the Cepheid method (Sakai et al. 1996, Ferrarese et al. 2000,
Sakai et al. 2004), but requires much less
observing time. Over the last few years, WFPC2 snapshot surveys have made
use of the TRGB method to obtain 10\% distances for about 100 galaxies.
Further significant progress is expected because the Advanced Camera for
Surveys (ACS) probes about 1.5 mag deeper in an equal integration
time and acquires a field double in area.

  Apart from 36 members of the Local Group, there are, so far, 310
galaxies with distance estimates less than 7 Mpc. Among these,
distances to 174 galaxies have been measured with an accuracy of 10\%
based on the Cepheid method (N=12), or the luminosity of the TRGB (N=162).
A list of these galaxies is presented in ``Catalog of Neighboring Galaxies''
(CNG) by Karachentsev et al. (2004). The remaining galaxies have only
rough distance estimates from the luminosity of their brightest
stars, the Tully-Fisher relation, or from their membership in the known
nearby groups. In this restricted distance range the TRGB method is most
effective because it provides accurate distances to galaxies of
all morphologies with minimal observational demands.
The superior angular resolution of HST has been a critical
factor: 95\% of the TRGB distances have been obtained with
HST during the last four years. The remaining 136 galaxies are
suitable targets for a SNAP survey with the Advanced Camera at HST.

  The radial velocity -- distance relation for nearby galaxies is shown
in Figure 1. Within 7 Mpc there are two substantial groups of galaxies:
around M~81 and Cen~A. The average distances of 3.7 Mpc, and 3.8 Mpc,
respectively, are very similar. Members of these two groups are shown
in Figure 1 as open circles. The solid line corresponds to the Hubble
relation with $H_0 = 68$ km s$^{-1}$ Mpc$^{-1}$, curved at small distances
because of the decelerating gravitational action of the Local Group
(Lynden-Bell 1981, Sandage 1986) assuming a total mass of $1.3\times
 10^{12} M_{\sun}$. The largest deviations from the Hubble regression
are seen to be related to the virial motions of galaxies inside the M81
and Cen~A groups. An ``S --wave'' feature brackets the distance regime
of the two dominant groups: nearer galaxies have velocities that
exceed the Hubble flow and more distant galaxies have lower velocities.
This pattern is the signature of infall toward massive attractors.

  Karachentsev \& Makarov (1996, 2001) found that the local expansion on
a scale of 5 Mpc is significantly anisotropic, described by the tensor
$H_{ij}$ with values (81$\pm$3) : (62$\pm$3) : (48$\pm$5) in km s$^{-1}$.
The minor axis of
the ellipsoid is directed towards the pole of the Local Supercluster, and
the major axis has an angle of (29$\pm5)\degr$ with respect to the direction
of the Virgo Cluster (shifted towards the Great Attractor position).
Later, Karachentsev et al. (2003) have used TRGB distances derived with
WFPC2 data to confirm the local velocity anisotropy. Some galaxies in
Figure 1 (like UGC~3755)
that lie well below the Hubble line are all found at high
supergalactic latitudes. Given the observed concentration of
nearby galaxies to the Supergalactic equatorial plane, the retardation
from Hubble flow seen in the high latitude galaxies is expected.
It follows naturally from numerical action models of large scale flows
(Shaya, Peebles, \& Tully 1995) where mass is assumed to be distributed
like the light of all the known galaxies (in a census that extends to
the Great Attractor).

  The accurate distances that have already been obtained with the HST
snap surveys provide tremendous new constraints for numerical action
modeling. It is to be appreciated that peculiar motions are best
constrained locally. A 10\% error at a Hubble distance of 300 km s$^{-1}$
translates
to an uncertainty in peculiar velocity of only 30 km s$^{-1}$ but the error
increases linearly with distance. Every accurate distance provides an
additional constraint on the local distribution of matter. The widest
possible coverage of the sky is desired.

  According to the results of N-body simulations (Governato et al. 1997,
Massio et al. 2005)
the dispersion in the motions of field galaxies and group centers
around the mean flow, $\sigma_v$, contains information on galaxy formation,
the density of matter, $\Omega_m$, and the relative importance of Dark Energy,
$\Omega_{\lambda}$. Early hints of a cold local Hubble flow ( Sandage et al.
1972, Tully 1982) have been confirmed by the results of HST TRGB programs
(Karachentsev et al. 2003c). Galaxies within 3 Mpc of the Local Group yield
a surprisingly low dispersion, $\sigma_v$ about 25 -- 30 km s$^{-1}$,
and the dispersion of the centroids of the nearest groups (Local, M~81,
IC~342, Scl, Centaurus~A, M~83, Canes Venatici~I groups) is
$\sim$25 km s$^{-1}$. Quiescence of the
local Hubble flow is a signature of a vacuum-dominated universe (Chernin 2001,
Baryshev et al. 2001). Cosmological simulations based
on the Cold Dark Matter (CDM) paradigm but without Dark Energy inevitably
produce models with local random motions that are too high.
Maccio et al. (2005) show that CDM models with Dark Energy can result in
low random motions in the density regime of the local neighborhood.
The earlier in history the transition from the dominance of attraction
to the dominance of repulsion, the lower the random motions expected
today due to the enhanced effect of adiabatic cooling with expansion.

\section{ HST ACS photometry}

  Our sample consists of 25 galaxies observed with the Advanced Camera
for Surveys (ACS) as part of an HST Cycle 12 snapshot project (\#9771).
We obtained 1200s F606W and 900s F814W images of each galaxy using
ACS/WFC with a CR-split of two.  The cosmic ray cleaned images (CRJ data
sets) were obtained from the STScI archive, having been processed
according to the standard ACS pipeline.

Stellar photometry was obtained using the ACS module of DOLPHOT (Dolphin
et al. in prep), using the recommended recipe and parameters.  In brief,
this involves the following steps.  First, pixels that are flagged as bad
or saturated in the data quality images were marked in the data images.
Second, pixel area maps were applied to restore the correct count rates.
Finally, the photometry was run.

In order to be reported, a star had to be recovered with S/N of at least
five in both filters, be relatively clean of bad pixels (such that the
DOLPHOT flags are zero) in both filters, and pass our goodness of fit
criteria ($\chi \le 2.5$ and $\vert sharp \vert \le 0.3$).

To estimate our photometric uncertainties and completeness, artificial
star tests were run on the KK230 and NGC247 fields, which represent the
most and least crowded images.  Completeness plots are shown in Figure
2.  We note that the plateau at $\sim 85$\% completeness is due to bad
pixels and cosmic rays.  We also show the magnitude errors as a function
of recovered magnitude in Figure 3.

CTE corrections were made according to ACS ISR03-09, and our zero points
and transformations were made according to Sirianni et al. (2005).  We
estimate the uncertainties in the calibration to be 0.05 to 0.10
magnitudes.

   We determined the TRGB using a Gaussian-smoothed $I$-band luminosity
function for red stars with colors $V-I$ within $\pm0\fm5$ of the mean
$\langle V-I \rangle$ expected for red giant branch stars. Following
Sakai et al. (1996), we applied a Sobel edge-detection filter.
The position of the TRGB was identified with the peak in the
filter response function.
According to Da Costa \& Armandroff (1990), for metal-poor systems the TRGB
is located at $M_I = -4.05$ mag. Ferrarese et al. (2000) calibrated the
zero point of the TRGB from galaxies with Cepheid distances and estimated
$M_I = -4\fm06 \pm0\fm07(random)\pm0.13(systematic)$. A new TRGB calibration,
$M_I = -4\fm04 \pm0\fm12$, was made by Bellazzini et al.(2001) based on
photometry and on a distance estimate from a detached eclipsing binary in the
Galactic globular cluster $ \omega$ Centauri. For this paper (as for
our previuos works with the HST data) we use $M_I = -4\fm05$.

\section{TRGB distances and integrated properties}

  ACS images of 25 observed galaxies are shown in Figure 4. The compass in
each field indicates the North and East directions. Usually our target
galaxies were centered on the middle of the ACS field, but for two
large targets (NGC~247, NGC~4605) the ACS position was shifted towards
the galaxy periphery to decrease a stellar crowding effect. In Figure 5
$I$ versus $(V-I)$ color magnitude diagrams (CMDs) for the 25 galaxies
are presented.

  A summary of the resulting distance moduli for the observed galaxies
is given in Table 1. Its columns contain: (1) galaxy name, (2) equatorial
coordinates, (3) radial velocity in km s$^{-1}$ in the Local Group (LG)
rest frame with the apex parameters from NASA Extragalactic Database (NED),
(4) position of the TRGB in apparent I-band magnitude,
(5) the total number of detected stars per image,
(6) Galactic extinction in the I-band from Schlegel et al. (1998),
(7) true distance modulus in mag, and (8) linear distance in Mpc.
Taking into account the typical uncertainty in the TRGB ($\sim0\fm15$),
as well as uncertainties of the HST photometry zero point ($\sim0\fm05$),
the aperture corrections ($\sim0\fm05$), and the crowding effects
($\sim0\fm06$) added quadratically, we estimate the uncertainty in the
derived distances to be about $8\%$.

  Some additional comments about the galaxy properties are briefly
discussed below. The galaxies are listed in order by increasing
Right Acsention.

  {\bf ESO 349-031= SDIG.} This dIr galaxy belongs to a
group (filament) in Sculptor. Its TRGB distance of 3.21~Mpc turns 
out to be 0.9 Mpc
less than a former estimate derived from the Tully-Fisher (TF)
relation by Karachentsev et al. (2003a). Judging from its distance and
radial velocity, SDIG is a companion to bright spiral galaxy NGC~7793.

  {\bf NGC 247.} Among the known galaxies in the Sculptor filament, this
spiral galaxy remained the last one without reliable distance estimate.
Dimensions of NGC 247 (27$\arcmin\times7\arcmin$) extend far beyond the
ACS field. We targeted the northern side of the galaxy at a distance
of 6$\farcm5$ from the nucleus which is relatively free from bright stellar
complexes. Nevertheless, about $1.7\times 10^5$ stars have been detected in
this field. The derived TRGB distance to NGC~247, 3.65 Mpc, agrees with
the distance estimate, 4.1 Mpc, obtained from the TF- relation
(Karachentsev et al. 2003a).

  {\bf KKH 6.} According to its radial velocity and distance, this dIr
galaxy is located at a far outskirt of the group around the
galaxy pair Maffei~1 and Maffei~2. The structure of this group is
strongly compromised by Galactic extinction.

  {\bf UGCA 86.} This Magellanic type irregular galaxy with a prominent
region of star formation (VII~Zw~9) at its southern side has been imaged
with WFPC2 by Karachentsev et al. (2003b), however, the former observations
yield only the lower limit of its distance, $D > 2.2$ Mpc. Our present
distance estimate, 2.96 Mpc, confirms the suggestion that
UGCA~86 is a companion to the giant spiral IC~342 situated at a distance
of 3.28 Mpc (Saha et al. 2002).

 {\bf UGCA 92.} This irregular galaxy with low radial velocity,
$V_{LG} = 89$ km s$^{-1}$, is situated in a zone of strong extinction.
Karachentsev et al. (1997) estimated its distance to be 1.8 Mpc based on the
luminosity of brightest blue stars. With the new TRGB distance, 3.01 Mpc,
the galaxy, as well as UGCA~86, turns out to be a physical member of
the group around IC~342. Note that 75 arcmin away from UGCA 92 there is
another galaxy, NGC 1569, of almost the same radial velocity $V_{LG} =
88$ km s$^{-1}$. Makarova \& Karachentsev (2003) carried out stellar
photometry of two HST/WFPC2 fields at a far periphery of NGC 1569 and
derived two possible distance estimates: 1.95$\pm$0.2 Mpc and
2.8$\pm$0.2 Mpc. The second estimates agrees better with the new distance
of UGCA 92.

 {\bf ESO 121-20 = KKs 19.} This very isolated dIr galaxy was detected
in the HI line by Huchtmeier et al. (2001). Its individual distance
estimate, 6.05 Mpc, is derived by us for the first time. With this
distance, the galaxy has a high hydrogen mass-to-luminosity ratio,
$M(HI)/L_B$ = 2.9 as well as high indicative mass-to-luminosity ratio,
$M_{25}/L_B$ = 18.5 in solar units that gives evidence of the existence
around ESO 121--20 of an extended HI envelope like around some other
isolated irregular galaxies: DDO~154 and NGC~3741 (Begum et al. 2005).

 {\bf KKH 37 = Mailyan 16.} This dIr galaxy of low surface
brightness is situated at the northern periphery of the IC~342 group.
The TRGB distance to KKH~37, 3.39 Mpc, is just the same as to IC~342
and NGC~2403. In this obscured region of sky there may be
other dwarf galaxies, undetected so far, which fill the gap between
the two
groups around these bright spiral galaxies.

 {\bf ESO 059-01.} This is an isolated irregular galaxy with a high
indicative mass-to-luminosity ratio, $M_{25}/L_B = 15.2 M_{\sun}/L_{\sun}$
at the TRGB distance $D$ = 4.57 Mpc.

 {\bf DDO 52 = UGC 4426.} This dIr galaxy is very isolated.
The nearest galaxy of normal luminosity, NGC~2683, is located about
2.5 Mpc away from DDO~52. The present case demonstrates to us that a galaxy
of 10 Mpc distance is reachable with ACS in a single orbit SNAP
observation.

 {\bf D564-08.} This dIr galaxy of low surface brightness from a list of
Taylor et al. (1996) is a probable distant companion to the giant spiral
galaxy NGC 2903. Its radial velocity, $V_{LG} = 365$ km s$^{-1}$
is surprisingly low for the distance of 8.67 Mpc, giving an
individual
Hubble ratio  $H_i = 42$ km s$^{-1}$ Mpc$^{-1}$. A similar low value,
$H_i = 39$ km s$^{-1}$ Mpc$^{-1}$, is also found for the
isolated galaxy DDO~52 discussed above.

 {\bf D634-03.} This is the most unusual object of low surface brightness
from the list by Taylor et al. (1996). Its low radial velocity,
$V_{LG} = 172$ km s$^{-1}$ measured by Schombert et al. (1997) was
later confirmed by Huchtmeier (personal communication),
who derived $V_{LG} = 173$ km s$^{-1}$, the
HI line width $W_{50} = 34$ km s$^{-1}$, and the HI flux 
$F = 0.25$~Jy~km~s$^{-1}$.
The RGB of D634-03 is not seen distinctly in its CM diagram. A probable
position of the TRGB seems to be near $I = 25\fm9$ which corresponds
to the galaxy distance 9.46~Mpc. With an apparent total magnitude of
$B = 17\fm5$, D634-03 has an absolute magnitude of $-$12.54, the hydrogen
mass-to-luminosity ratio 0.37 $M_{\sun}/L_{\sun}$, and the total (indicative)
mass-to-luminosity ratio 10 $M_{\sun}/L_{\sun}$. An individual Hubble ratio
of this isolated galaxy is only 18 km s$^{-1}$ Mpc$^{-1}$ indicating 
a high peculiar velocity.

 {\bf D565-06.} One more galaxy of low surface brightness from
Taylor et al. (1996). Its low radial velocity, $V_{LG} = 386$ km s$^{-1}$,
contrasts with the large TRGB distance, 9.08 Mpc. Based on this
distance, D565-06 is a companion to NGC 2903, which has
$V_{LG} = 442$ km s$^{-1}$ and a distance of 8.9 Mpc from the 
luminosity of its brightest stars.

 {\bf IKN.} This galaxy of very low surface brightness is a new dwarf
spheroidal member of the M 81 group. IKN is situated 1$\arcmin$ south from a
bright star, mimicking the star reflex on POSS-II plates. Our ACS
photometry reveals $2\times 10^4$ stars seen in both filters. Most of them
belong to RGB population. Near the center of IKN, we found a candidate
globular cluster marked in Fig. 4 by a circle. From its dimension
(2$\farcm$7) and apparent magnitude, $B_T = 16.0$, IKN resembles another dSph
member of this group, F8D1 observed with WFPC2 by Caldwell et al. (1998).

  {\bf HS 117.} This dwarf system was found by Huchtmeier \& Skillman (1998),
who detected it in the HI line with $V_h = -37$ km s$^{-1}$, i.e. in the
range of
Galactic hydrogen emission. The CM diagram of HS 117 on Fig. 5  exhibits
a distinct sequence of RGB stars as well as some number of bluish stars.
Based on its optical and HI properties, HS 117 may be classified as
a dwarf galaxy of the transition dIr/dSph type.

  {\bf NGC 4068.} This blue dwarf galaxy belongs to the Canes Venatici~I
(CVn~I) cloud. Makarova et al. (1997) estimated its distance via the brightest
stars to be 0.9~Mpc greater than the new distance of 4.31~Mpc to 
NGC~4068 derived by us from the TRGB.

 {\bf NGC 4163.} This is another blue and compact galaxy in the
CVn~I cloud. Its old distance, 3.6 Mpc, was estimated by Tikhonov \& Karachentsev
(1998) from the brightest stars.  Our new estimate of 2.96~Mpc from
the TRGB places
NGC~ 4163 at the nearby edge of the CVn~I cloud which makes it the
second nearest BCD galaxy.

 {\bf UGC 7242 = KKH 77.} A bright star projected into the northern
side of this dIr galaxy renders stellar photometry of UGC~7242 to be
incomplete. The galaxy is situated at the far outskirt of the M~81 group,
behind the main body of the group.

 {\bf IC 779 = UGC 7369.} This is a lenticular galaxy with a 
semi-stellar
nucleus. Despite its low radial velocity, $V_{LG} = 198$ km s$^{-1}$,
IC~779
does not look to be a nearby object. For this reason it was excluded from the
Local Volume galaxies (see Table 3 in CNG, Karachentsev et al. 2004).
The distribution of the resolved
red stars over the ACS area reveals that most are 
concentrated within the main optical boundary of the galaxy. 
The CM diagram of IC 779 in Fig.5 shows what might be an upper part
of the RGB
with the TRGB position at $I \sim 26\fm3$. Alternatively,  these 
stars might be part of the asymptotic giant branch (AGB).  From 
its position on the sky, it is suspected that this galaxy is a 
member of the Coma~I group.  Assuming we are picking up the TRGB, 
the derived distance to IC~779 is 11.6 Mpc.  The plausible 
association with the Coma~I group would place the galaxy at 16~Mpc
(Tonry et al. 2001) whence the TRGB would be lost at the 
$I \sim 27$ limit of our observations.  This ambiguity created by
AGB populations demonstrates
the breakdown of the TRGB method near the photometric limit.
In the case of single orbit observations with ACS of galaxies
with young populations, the limit of
reliable TRGB measures is $\sim 10$~Mpc.

 {\bf NGC 4605.} This bright Sdm galaxy is situated in the general field
between the M81 group and CVn~I cloud. Our ACS observations were pointed on
the eastern side of NGC~4605 about 1$\farcm$5 far from its star-like nucleus.
In the ACS field we discover more than $10^5$ stars seen in both $V$ and $I$
filters. An overwhelming majority of them belong to the RGB population
with TRGB position at $I = 24\fm67$, yielding the galaxy distance to
be 5.47 Mpc.

 {\bf HIPASS J1247-77.} This is a dIr galaxy found in the HIPASS
blind survey (Kilborn et al. 2002). The galaxy is situated in the
zone of strong extinction, $E(B-V) = 0.75$, and its  CMD is heavily
contaminated by Milky Way stars. The CM diagram derived by us for the
stars located only within the optical boundary of the galaxy (see Fig. 5)
reveals the red giant branch population with the TRGB position at
$I = 24\fm9$, corresponding to the distance of 3.16 Mpc. In our list,
HIPASS J1247-77 is the only galaxy originally discovered in the HI line,
but its global parameters: $M(HI)/L_B = 0.20$ and $M_{25}/L_B = 1.0$ in
solar units do not distinguish it from other dIr galaxies.

 {\bf UGC 8215.} This dIr galaxy in the CVn~I cloud was resolved into
stars for the first time by Makarova et al. (1997), who estimated its
distance to be 5.6 Mpc via the brightest stars. The CM diagram derived
from our ACS data yields the TRGB distance of 4.55 Mpc which does not
change its membership in the CVn~I cloud.

 {\bf UGC 8638.} This is another compact irregular galaxy in the direction of
the CVn~I cloud. Because of the presence in UGC 8638 of bright compact stellar
associations, Makarova et al. (1998) were able to determine only the
lower limit to its distance, $D > 2.3$ Mpc. Our ACS photometry of
UGC 8638 yields the TRGB distance 4.27 Mpc, typical for the CVn~I cloud.

 {\bf KK 230 = KKR 3.} This dIr galaxy of very low surface brightness
is one of the faintest ones among the known galaxies of the general field.
Our ACS photometry yields for KK~230 the TRGB distance of 1.92 Mpc
in agreement with a previous TRGB estimate, 1.90 Mpc, obtained by
Grebel (personal communication) from observations with the 10-m Keck
telescope. We also performed surface photometry of the galaxy based
on its ACS images and derived the galaxy magnitude $I(R < 40\arcsec$)
= 15$\fm$6$\pm0\fm15$, the integrated color $(V - I) = 0.90$
inside the same radius $R$, the central $I$ band surface brightness
$22.9\pm0.2^m/\sq\arcsec$, and
the exponential profile scale $(13.4\pm0.1)\arcsec$. Assuming for KK~230 a
typical color $(B - V) = 0.50$, we estimate its integrated blue magnitude
to be $B = 17\fm0\pm0\fm25$, yielding the absolute magnitude $M_B = -9.47$.
Then, the hydrogen mass-to-luminosity ratio and the indicative mass-to-
luminosity ratio are 2.3 and 3.3 in solar units, respectively.

  {\bf IC 4662.} This is an isolated dwarf galaxy of high surface
brightness with star formation complexes especially prominent at the
northern galaxy side. We measured for IC~4662 the TRGB distance to be
2.44 Mpc. That makes this galaxy the nearest known representative of
BCD objects. It is amazing that this bright ($ B = 11\fm74$) galaxy
with the radial velocity of $ V_{LG} = 145$ km s$^{-1}$ was never
resolved into stars before.

 {\bf KK 246 = ESO 461-036.} The irregular galaxy of low surface
brightness is the most isolated system in the Local Volume. Being at
the TRGB distance of 7.83 Mpc, KK~246 is situated just at the edge of
the Local Void described by Tully (1988). Nevertheless, the global
parameters of KK~246: $M(HI)/L_B = 2.4 M_{\sun}/L_{\sun}$ and
$M_{25}/L_B = 4.7 M_{\sun}/L_{\sun}$
do not distinguish KK~246 from other dIrs situated
in the known nearby groups.

\section { Peculiar velocities of nearby galaxies}

 A significant part of the scatter of nearby galaxies in the Hubble
diagram (Fig. 1) is caused by the virial motions in groups. According
to Karachentsev (2005), the mean-square virial velocity of galaxies along
the line of sight in 9 nearby groups ranges from 54 km s$^{-1}$
 (IC~342
group, Sculptor filament) to 105 km s$^{-1}$  ( Cen~A group) with the median
$\sigma_v = 71$ km s$^{-1}$. To study properties of peculiar
velocities of the
field galaxies themselves, we refined the sample of LV galaxies to
exclude group members, keeping only galaxies with the tidal index  
$TI < 0$.
By the definition of $TI$ (Karachentsev et al. 2004), such a galaxy has a
crossing time with respect to its significant neighbor larger than the
cosmic expansion time, $H^{-1}$. For each of these 110 field galaxies, we
determined an individual Hubble ratio, $H_i = (V_{LG})/D$. The distribution
of this parameter for the LV galaxies is presented in the left
panel of Fig. 6. Typical errors on radial velocities ( $< 5$ km s$^{-1}$)
do not appreciably distort this distribution.
The bulk of galaxies on the histogram follow a Gaussian
distribution with the mean  $<H> = 68$ km s$^{-1}$ Mpc$^{-1}$ and the
standard deviation
$\sigma(H) = 15$ km s$^{-1}$ Mpc$^{-1}$ that is 2.2 times larger than
the expected scatter
caused by 10\% errors in distance measurements. On the left shoulder of
the histogram, we recognize an excess number of galaxies with $H =
(8 - 32)$ km s$^{-1}$ Mpc$^{-1}$.

  Adopting the mean Hubble parameter for the LV galaxies to be
68 km s$^{-1}$ Mpc$^{-1}$,
we obtain a distribution of the peculiar velocities of nearby
isolated galaxies, 
$V_{pec} = V_{LG} - <H> D$, seen on the right panel of Fig. 6.
Again, the number galaxy distribution can be described by a Gaussian law
with $\sigma(V_{pec}) = 63$ km s$^{-1}$ and an asymmetric tail extended
far towards negative values.

  Apart from 110 isolated LV galaxies with accurate distance estimates
via cepheids or TRGB, we added to histograms in Fig. 6 five galaxies,
which were listed in Table 3 of CNG as distant
objects with low radial velocities. Distances to these galaxies were
estimated by less precise methods (from surface brightness fluctuations,
SBF, Tully-Fisher relation, TF, or luminosity of the brightest stars, BS).
Distance estimates for these galaxies are greater than 10 Mpc, but
their radial velocities satisfy a condition $V_{LG} < 500$ km s$^{-1}$. Basic
parameters of these are listed in Table 2. Besides NGC~1400, situated in
the Eridanus group, the others look to be
quite isolated objects. They all are placed in Fig. 6
as the extreme cases with  $H < 32$ km s$^{-1}$ Mpc$^{-1}$
or $V_{pec} < -500$ km s$^{-1}$.

  Different reasons can cause the observed asymmetric distribution
of galaxies on their peculiar velocities with a skewness toward negative
values. An anisotropy of the Local Hubble flow found by Karachentsev \&
Makarov (1996, 2001) is among them. The main feature of this phenomena
is the low local Hubble parameter value seen in directions toward the
poles of the Local Supercluster. Such an effect is expected because of
gravitation decelerating of galaxies ofset from the main (most
dense) plane of the Supercluster. The character of the local anisotropy is
seen in Fig. 7, which displays peculiar velocities of galaxies versus
the elevation above  the Supergalactic plane as an angle SGB
(left panel) and in megaparsecs as SGZ (right panel). In spite of 
a considerable
scatter of galaxies on the diagrams (caused, in particular, by a residual
anisotropy of the Hubble flow within the Supergalactic plane), the trends
of $V_{pec}$ with SGB and SGZ are clearly seen. The regression
$< V_{pec}|SGB>$ has a slope $(-2.5\pm0.5$) km s$^{-1} deg^{-1}$, and the
regression $< V_{pec}|SGZ>$ is characterized by a slope $(-43\pm4)$
km s$^{-1}$ Mpc$^{-1}$. The last
quantity is of order of the Hubble parameter, meaning that the local
anisotropy on Z-coordinate,  $d|V_{pec}|/dZ = -0.6 <H>$, turns out to be
rather significant.

  The distribution on the sky of nearby galaxies with large negative
peculiar velocities looks very clumpy. Among ten galaxies with moderate
peculiar velocities, [$-$200, $-$500] km s$^{-1}$,
seven: UGC~3755, DDO~47, KK~65, UGC~4115, D564-08, D634-03,
and D565-06 are situated in a filament aligned along 25$\degr$ in SGB
in the Gemini -- Cancer constellations. Moreover, six of the seven galaxies
with $V_{pec} < -500$ km s$^{-1}$ are concentrated within
a small ring of radius 5$\degr$ in the Coma constellation just at
the Supergalactic equator.  Four of these 6 are superposed on the
Coma~I group, a region of 5 x 5 $\degr$ around NGC 4494 identified 
as group 14-1 in the Nearby Galaxies Catalog (Tully 1987).
This tight group of early-type galaxies at a distance of 16~Mpc
(Tonry et al. 2001) has a mean heliocentric velocity of 
902~km~s$^{-1}$.  Its dispersion of 283~km~s$^{-1}$ has the
consequence that several of its members scatter into our 
$V_{LG} < 550$~km~s$^{-1}$ sample.

  Figure 8 presents a distribution of the distances
and peculiar velocities of nearby galaxies. Two straight lines, 
symmetric with respect to
$V_{pec} = 0$, indicate a sector of 10\% errors in galaxy distance estimates.
It is seen that about 2/3 of the galaxies are situated outside this sector.
The inclined solid line indicates a zone of observational selection
corresponding to  $V_{LG} = 550$ km s$^{-1}$ : galaxies without
individual distance estimates were included into the CNG
only if their corrected radial velocities do not exceed 550 km s$^{-1}$.
Therefore, one may expect the existence of populations of other galaxies,
like NGC~925, which have $V_{LG} > 550$ km s$^{-1}$, but $D < 10$ Mpc.

 \section { Concluding remarks}

  In the Catalog of 451 LV galaxies (CNG), there are
only 214 galaxies with precise distance estimates from luminosities of
cepheids and the TRGB, including our present ACS observations of 25 galaxies.
Among these, we find a few galaxies whose radial velocities
differ from the expectations of a homogeneous isotropic
Hubble flow. Considering the sample of 110 LV galaxies that have accurate
distance estimates and are situated outside known nearby groups, we
derive the mean value of a local Hubble constant to be 68
km s$^{-1}$ Mpc$^{-1}$ and the 1D peculiar velocity dispersion
of $\sigma_v = 63$ km s$^{-1}$. Part of this
velocity dispersion is caused by anisotropy of the local Hubble flow
since the rate of expansion along the Local Supercluster poles is about
half that in the LSC plane.

  The all-sky distribution of nearby galaxies with high peculiar velocities
is unusual. A majority of galaxies with  
$-500 < V_{pec} < -200$~km~s$^{-1}$
are situated in Gemini -- Cancer constellations, forming a filament of
$\sim25\degr$ length along an SGB meridian. 
This area lies in the region of the 'Leo Spur' in the nomenclature
of the Nearby Galaxies Catalog (Tully 1987), the most pronounced
part of the 'Local Velocity Anomaly' (Tully, Shaya, and Pierce 
1992).  In the model presented by those authors, the Leo Spur is
converging on the filament in which we live, the two structures
drawn together by their mutual attraction.
The Gemini -- Cancer region is located within the zone of
a systematic ``blind'' survey in the HI line at Arecibo
(the project ALFALFA, Giovanelli et al. 2005). This deep survey likely will
discovery more objects within the region of the Local Velocity
Anomaly.

If one looks around the sky for all galaxies with 
$V_{pec} < -500$ km s$^{-1}$, excluding the well-known cases
that lie in the Virgo and Fornax clusters, one currently finds
7 examples.  Four of these (NGC 4150, UGC 7131, IC 779, KK 127) 
lie in the tight knot of early type galaxies around NGC 4494 commonly
called the Coma~I group.  One (NGC 1400) is in a similar knot of 
early type galaxies around NGC 1407 (Gould 1993).  Another
(UGC 7321) can probably be attributed to backside infall into
the Virgo Cluster.  The remainder (UGC 6782) may likewise be
attributed to Virgo infall or may be assigned a large peculiar 
velocity due to a falacious distance.  

Perhaps it is surprising that there are so {\it few} systems with
large anomalous velocities.  Those few are in very small sectors 
of the sky and almost all are projected onto the nearest important
knots of early-type systems; regions of high inferred mass
concentration.  The intermediate amplitude peculiar velocities 
in the Gemini - Cancer constellations appear to be reflections of
large-scale streaming; in this case, probably the result of
the attraction between the nearest pair of filaments, each with 
masses in the range of $10^{14} M_{\odot}$ (Tully et al. 1992).
Otherwise, galaxy velocities are remarkably quiet about the local 
expansion, with rms random motions only $\sim 60$~km~s$^{-1}$.

  \acknowledgements{
Support associated with HST program GO--9771 was provided by NASA 
through a grant from the
Space Telescope Science Institute, which is operated by the Association of
Universities for Research in Astronomy, Inc., under NASA contract
NAS5--26555.
This work was also supported by RFFI grant 04--02--16115.}

{}

\newpage
\figcaption{The radial velocity - distance relation for nearby
	    galaxies with accurate distance estimates. Members of
	    the M~81 group and Cen~A group at $\sim$3.8 Mpc are shown as
	    open circles. The galaxies in other nearby groups are
	    indicated as open squares. The field galaxies with the
	    tidal index TI < 0 are presented as filled circles.
	    The solid line corresponds to the Hubble relation with
	    $H_0 = 68$ km s$^{-1}$ Mpc$^{-1}$, curved at small distances
	    because of the decelerating gravitational action of the Local
	    Group assuming a total mass of $1.3\times 10^{12} M_{\sun}$.}

\figcaption{Completeness as a function of magnitude for the least crowded
	    (KK230) and the most crowded (NGC247) objects.}

\figcaption{Mean error as a function of magnitude for KK230 and NGC247.}

\figcaption{ACS images of 25 nearby galaxies produced by two 600 second
	    exposures in F606W (``V'').}

\figcaption{ACS color-magnitude diagrams for 25 nearby galaxies.
	    For the small-angular-size galaxy HIPASS1247-77
	    the CMD is given for the stars inside its optical boundary
	    only.}

\figcaption{Number of nearby field galaxies as a function of individual
	    Hubble parameter (left panel) and peculiar velocity,
	    $V_{pec} = V_{LG} - 68 D$ (right panel).}

\figcaption{Peculiar velocity as a function of the Supergalactic latitude
	    (left panel) and distance from the Supergalactic plane
	    (right panel).}

\figcaption{Peculiar velocity of nearby field galaxies as a function
	    of distance.}

\newpage

\renewcommand{\arraystretch}{.6}
\begin{deluxetable}{lcrrrcrr}
\tablewidth{0pc}
\tablecaption{ New distances to nearby galaxies}
\tablehead{
\colhead{Name}&
\colhead{RA(J2000.0)Dec}&
\colhead{$V_{LG}$}&
\colhead{I(TRGB)}&
\colhead{$N_{**}$}&
\colhead{$A_I$}&
\colhead{$(m-M)_0$}&
\colhead{$D$}\\
& & \colhead{km s$^{-1}$}&
\colhead{mag}&
&\colhead{mag}&
\colhead{mag}&
\colhead{Mpc}}
\startdata
E349-031,SDIG & 000813.3$-$343442 &   216 &  23.50&  13 990 &  0.02 & 27.53 &  3.21 \\
N247          & 004708.3$-$204536 &   215 &  23.80& 174 560 &  0.04 & 27.81 &  3.65  \\
KKH6          & 013451.6$+$520530 &   270 &  24.49&   8 150 &  0.68 & 27.86 &  3.73 \\
UA86          & 035949.5$+$670731 &   275 &  25.14&  69 260 &  1.83 & 27.36 &  2.96 \\
UA92          & 043200.3$+$633650 &    89 &  24.88&  30 840 &  1.54 & 27.39 &  3.01 \\
	      &                  &       &       &        &       &       &
  \\
E121-20,KKs19 & 061554.5$-$574335 &   309 &  24.94&   9 350 &  0.08 & 28.91 &  6.05 \\
KKH37         & 064745.8$+$800726 &   204 &  23.75&  10 900 &  0.15 & 27.65 &  3.39 \\
E059-01       & 073119.3$-$681110 &   245 &  24.53&  42 050 &  0.28 & 28.30 &  4.57 \\
DDO52         & 082828.5$+$415124 &   398 &  26.08&   9 950 &  0.07 & 30.06 & 10.28 \\
D564-08       & 090254.0$+$200431 &   365 &  25.69&   3 070 &  0.05 & 29.69 &  8.67 \\
	      &                 &       &       &        &       &       &
 \\
D634-03       & 090853.5$+$143455 &   173 &  25.90&   1 860 &  0.07 & 29.90 &  9.46 \\
D565-06       & 091929.4$+$213612 &   386 &  25.82&   2 670 &  0.08 & 29.79 &  9.08 \\
IKN           & 100805.9$+$682357 &  $-$  &  23.94&  19 230 &  0.27 & 27.87 &  3.75 \\
HS117         & 102125.2$+$710658 &   116 &  24.16&   5 890 &  0.22 & 27.99 &  3.96 \\
N4068         & 120402.4$+$523519 &   290 &  24.16&  75 830 &  0.09 & 28.17 &  4.31 \\
	      &                 &       &       &        &       &       &
 \\
N4163         & 121208.9$+$361010 &   164 &  23.35&  68 180 &  0.04 & 27.36 &  2.96 \\
U7242,KKH77   & 121407.4$+$660532 &   213 &  24.66&  28 240 &  0.04 & 28.67 &  5.42 \\
IC779,U7369   & 121938.8$+$295259 &   198 &  26.3:&   2 100 &  0.03 & 30.32 & 11.6 \\
N4605         & 124000.3$+$613629 &   276 &  24.67&  80 930 &  0.03 & 28.69 &  5.47 \\
HIPASSJ1247-77& 124732.6$-$773501 &   155 &  24.9 &   2 560 &  1.44 & 27.50 &  3.16 \\
	      &                 &       &       &        &       &       &
 \\
U8215         & 130803.6$+$464941 &   297 &  24.26&   7 780 &  0.02 & 28.29 &  4.55 \\
U8638         & 133919.4$+$244633 &   273 &  24.13&  29 850 &  0.03 & 28.15 &  4.27 \\
KK230         & 140710.7$+$350337 &   126 &  22.40&   4 040 &  0.03 & 26.42 &  1.92 \\
IC4662        & 174706.3$-$643825 &   145 &  23.02& 167 770 &  0.13 & 26.94 &  2.44 \\
KK246,E461-36 & 200357.4$-$314054 &   401 &  26.0:&   2 990 &  0.58 & 29.47 &  7.83 \\

\enddata
\end{deluxetable}

\begin{deluxetable}{lcccrll}
\tablewidth{0pc}
\tablecaption{List of distant galaxies with low velocities}
\tablehead{
\colhead{Name}&
\colhead{RA(J2000.0)Dec}&
\colhead{$V_{LG}$}&
\colhead{$D$}&
\colhead{$V_{pec}$}&
\colhead{Distance reference}\\
& & \colhead{km s$^{-1}$}&
\colhead{Mpc}&
\colhead{km s$^{-1}$} & &}
\startdata
N1400 & 033930.6$-$184115 & 485 &  26.4 & $-$1310 &  SBF, Tonry et al. (2001)
    \\
U6782 & 114857.3$+$235016 & 452 &  14.0 & $-$500  &  BS, Makarova et al. (1998)  \\
U7131 & 120911.8$+$305425 & 226 &  14.0 & $-$726  &  BS, Makarova et al. (1998)  \\
N4150 & 121033.6$+$302406 & 198 &  13.7 & $-$733  &  SBF, Tonry et al. (2001)
   \\
KK127 & 121322.7$+$295518 & 105 &  13.0 & $-$779  &  SB vs. Lumin., present paper\\
U7321 & 121734.1$+$223222 & 345 &  20.0 & $-$1015 &  IR-TF, present paper  \\

\enddata
\end{deluxetable}

\end{document}